\documentclass[twocolumn,superscriptaddress, floatfix, showpacs, aps, prl, 10pt]{revtex4-2}
\usepackage{graphicx}
\usepackage{amsmath}
\usepackage[colorlinks=true, citecolor=black, urlcolor=black]{hyperref}
\usepackage{amssymb}
\usepackage{bm}
\usepackage{color}
\usepackage{array}
\usepackage{booktabs}
\usepackage{multirow}
\usepackage{braket}
\usepackage{xcolor}
\usepackage{soul}

\begin{document}

\title{Saturating $T^2$ resistivity in iron-based superconductors }
\author{M. Lourdes Amig\'o}
\address{Instituto de Nanociencia y Nanotecnolog\'ia CNEA-CONICET, Argentina}
\author{Jorge I. Facio}
\address{Instituto de Nanociencia y Nanotecnolog\'ia CNEA-CONICET, Argentina}
\address{Instituto Balseiro, CNEA, 8400 Bariloche, Argentina}
\date{\today}

\begin{abstract}
Many correlated metals exhibit a crossover from low-temperature \( \rho \propto T^{2} \) behavior to a slower increase of the resistivity at elevated temperatures. Distinguishing phonon scattering from electron--electron correlations as the origin of this crossover is experimentally challenging. We introduce a simple phenomenological framework that isolates the contribution responsible for the high-temperature saturating tendency. This approach reveals a clear contrast between systems well known for the predominance of electron--phonon scattering and several iron-based superconductors. For the latter, the conductivity can be expressed as the sum of a Fermi-liquid-like term (\(\propto 1/T^{2}\)) and a temperature-independent component, which together describe \(\rho(T)\) up to \(\sim250\)\,K in LiFeAs, KFe$_2$As$_2$, and KCaFe$_4$As$_4$. 
In FeSe, the same approach identifies a similar regime above the nematic transition, with a pronounced change toward nearly linear-in-$T$ resistivity below it. 
We discuss these findings in the broader context of coherence and electronic correlations in Hund’s metals.
\end{abstract}

\maketitle

\setcounter{secnumdepth}{3}

\section{Introduction}

In the semiclassical picture of simple metals, the electrical conductivity $\sigma$ decreases monotonically with temperature $T$ as scattering events become more frequent. 
For instance, at low $T$, the electron-electron (e-e) scattering rate in a Fermi liquid increases as $\Gamma_{\mathrm{FL}} \propto T^2$, while electron-phonon (e-ph) backscattering scales as $T^5$.
There are, however, materials that behave as conventional metals at low $T$ but, at higher temperatures, exhibit a conductivity that either ceases to decrease or does so much more slowly over an extended temperature range~\cite{ioffe1960non,mott1972conduction,PhysRevB.24.7404,RevModPhys.75.1085,hussey2004universality}.
Salient examples include systems dominated by strong e-ph scattering~\cite{PhysRev.181.1127,PhysRevLett.36.1084,PhysRevLett.38.782} as well as correlated systems undergoing a coherence-incoherence crossover~\cite{haule2009coherence,georges2013strong}.
Identifying the mechanism driving the deviations from the low-temperature regime is an experimental challenge, largely due to the lack of a universal parametrization for the high-temperature e-e and e-ph resistivities.

Exploiting the additive nature of the conductivity associated with a specific scattering mechanism, we propose that the response dominating these deviations can be decomposed as
\begin{equation}
\sigma = \sigma_i + \sigma_s,
\label{eq_sigma_s}
\end{equation}
where $\sigma_i$ captures the conventional metallic behavior that dominates at low temperatures, while $\sigma_s$ represents the contributions responsible for the high-temperature saturating tendency.
We demonstrate that this additive decomposition is effective for analyzing both systems dominated by e-ph scattering—consistent with theoretical derivations~\cite{PhysRevLett.122.026602,werman2017non}—and systems governed by strong e-e correlations.
Crucially, the distinct temperature dependence of the extracted $\sigma_i$ allows for a clear distinction between these cases.

Based on data from several experimental studies~\cite{morozov2010single,PhysRevB.85.060503,PhysRevB.106.054508,zhang2019LiFeAs,nakajima2014strong,hardy2013evidence,10.1143/JPSJ.78.063702,doi:10.1143/JPSJ.79.124713,fang2015KFe2As2,PhysRevB.94.064501,PhysRevB.96.140501,singh2018CaKFe4As4,pyon2019CaKFe4As4,PhysRevB.110.224508,tanatar2016origin,kasahara2014field,watson2015emergence,zhou2021FeSe}, we show that for selected iron-based superconductors, this procedure yields an excellent description of the resistivity in a remarkably wide temperature range, identifying $\sigma_i$ with a Fermi-liquid-like term ($\propto 1/\Gamma_{\mathrm{FL}}$) and $\sigma_s$ with a nearly temperature-independent contribution. 
This leads to a distinct  ``saturating $T^2$" regime, which extends up to $\sim$250\,K in LiFeAs, KFe$_2$As$_2$, and KCaFe$_4$As$_4$.
In this regime, the deviations from the $T^2$ resistivity as the temperature increases are identified as intrinsic to the e-e scattering mechanism, rather than due to phonon onset.
Our analysis supports interpreting $\sigma_s$ as an intrinsic,
material-dependent parameter, analogous to the prefactor of the
low-temperature $T^2$ resistivity. We propose that this contribution
effectively reflects transport by short-lived electronic excitations at elevated temperatures.

The compounds discussed above preserve tetragonal symmetry down to the superconducting phase, and the saturating $T^2$ regime evolves smoothly into the low-temperature Fermi-liquid regime.
In FeSe, which in contrast undergoes a nematic transition at $T_s = 90\,$K~\cite{mcqueen-phase-trans,PhysRevLett.114.027001,watson2015emergence,shimojima2014lifting,fernandes2014drives,hosoi2016nematic}, our analysis reveals a sharp transition between a saturating $T^2$ behavior above $T_s$ (similar to the other compounds) and a nearly linear-in-$T$ dependence below $T_s$.
Our findings suggest a common incoherent transport regime in the high-temperature tetragonal phase of all these systems, with the recovery of a coherent Fermi liquid and the onset of nematic order representing two distinct low-temperature pathways by which this state is resolved.

This paper is organized as follows. In Sec.~\ref{sec:saturation}, we describe how Matthiessen's rule, combined with a phenomenological decomposition of the conductivity, leads to a modified parallel-resistor formula, which we first use to contrast LiFeAs, dominated by strong e-e scattering, with Nb, dominated by strong e-ph scattering. In Sec.~\ref{sec:other}, we apply the same analysis to KFe$_2$As$_2$, KCaFe$_4$As$_4$, and FeSe. In Sec.~\ref{sec:discussion}, we discuss our findings in terms of e-e versus e-ph scattering and the coherence-incoherence crossover in Hund's metals. Finally, we summarize our conclusions in Sec.~\ref{sec:conclusions}.

\section{Different types of resistivity saturation}
\label{sec:saturation}
\begin{figure}[t]
    \centering
    \includegraphics[width=1\columnwidth]{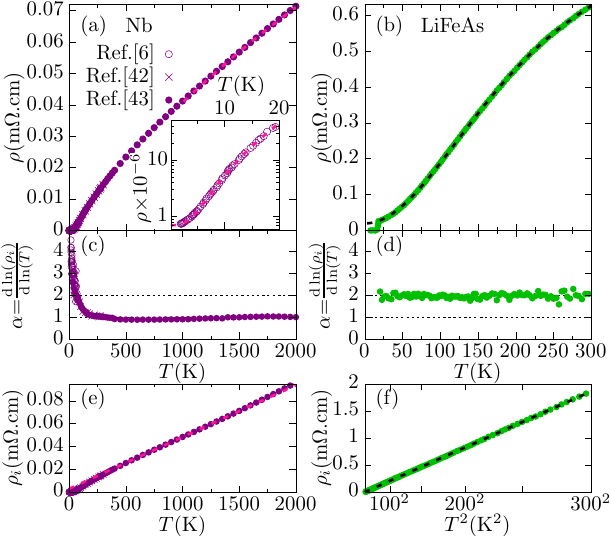}
	\caption{(a,b) Temperature dependence of the resistivity of Nb and LiFeAs, respectively. The data were digitized from Refs.\cite{PhysRev.181.1127,allen1986dc,abraham1972resistivite,morozov2010single}. The dotted line represents a fit to Eq.~\ref{eq_prf2}. Inset: low-temperature detail of the Nb fit. 
(c,d) For each compound, the temperature dependence of the logarithmic derivative of $\rho_i$, defined as $\alpha(T) = \frac{d\ln(\rho_i)}{d\ln(T)}$, where $\rho_i$ is obtained from the experimental data using Eq.~\ref{eq_prf2}. 
(e,f) $\rho_i$ as a function of $T$ for Nb and of $T^2$ for LiFeAs. The dotted line is a linear fit to the displayed data.}
    \label{fig:Nb_vs_Fe}
\end{figure}

We begin by contrasting $\rho(T)$ in two representative systems: Nb, a metal where transport is dominated by e-ph scattering~\cite{PhysRevB.54.16487}, and LiFeAs, a system that, like other iron-based superconductors, displays significant electronic correlations, evident in the large effective masses and bandwidth renormalizations~\cite{PhysRevLett.105.067002,PhysRevLett.108.047002,PhysRevB.88.144518,PhysRevB.93.085137, PhysRevB.104.125130}.  
Figs.~\ref{fig:Nb_vs_Fe}(a),(b) show the resistivity of both systems, with data taken from~\cite{PhysRev.181.1127,allen1986dc,abraham1972resistivite} and~\cite{morozov2010single}.
In both cases, the resistivity exhibits a power-law dependence at low temperatures, which gradually evolves into a sublinear-in-$T$ behavior. 

It is of interest to develop a phenomenological description of the emergence of this behavior without assuming the specific scattering mechanism responsible for it.
To this end, we build on a transport framework based on low-energy quasiparticles with finite lifetimes, as introduced in Ref.~\cite{pickem2021resistivity}.
In this approach, the conductivity is evaluated within linear response, assuming an energy- and momentum-independent scattering rate \(\Gamma\) and neglecting vertex corrections, and reads
\begin{equation}
	\sigma(T) = \frac{e^2}{h}\,\frac{K(T,\Gamma)}{\Gamma},
\label{eq_sigma}
\end{equation}
where \(K(T,\Gamma)\) is a transport kernel determined by the electronic structure, including quasiparticle weight and lifetime effects.
Its explicit form, discussed in Ref.~\cite{pickem2021resistivity}, is not essential for the phenomenology discussed here.

In the presence of distinct independent scattering mechanisms, 
the total scattering rate follows from the sum of the corresponding scattering rates (Matthiessen’s rule). 
An important consideration we make is that, at high temperatures, the resistivity is primarily limited by one mechanism. 
We denote the scattering rate of this mechanism as $\Gamma_{\text{HT}}$ and group all remaining contributions into $\Gamma_{\text{R}}$~\footnote{The subscript HT identifies the mechanism that dominates at high temperatures, but $\Gamma_{\text{HT}}$ refers to the scattering rate of that mechanism at \textit{all} temperatures.}. 
Thus, 
\begin{equation}
    \Gamma(T) = \Gamma_{\text{R}}(T) + \Gamma_{\text{HT}}(T).
    \label{eq_scarate}
\end{equation}
Using Eqs.~\ref{eq_sigma} and~\ref{eq_scarate}, we obtain
\begin{equation}
    \big(\rho - \rho_{\text{R}}\big)^{-1} = \rho_{\text{HT}}^{-1},
    \label{eq_prf}
\end{equation}
where $\rho$$=$$\sigma^{-1}$, $\rho_{\text{R}}$$=$$\Gamma_{\text{R}}/(\frac{e^2}{h} K)$ and $\rho_{\text{HT}}$$=$$\Gamma_{\text{HT}}/(\frac{e^2}{h} K)$.  

The right-hand side of Eq.~\ref{eq_prf} represents the conductivity limited by the mechanism responsible for the sublinear-in-$T$ behavior, denoted $\sigma_{\text{HT}}=\rho^{-1}_{HT}~$\footnote{As defined, $\rho_{\text{HT}}$ depends on $K$, which in turn depends on the total scattering rate $\Gamma$.  $\rho_{\text{HT}}$ can be directly interpreted as the resistivity of the mechanism labeled HT only when the contribution of other mechanisms is weak or when $K$ depends weakly on $\Gamma$.}.
In this context, we consider the possibility that $\sigma_{\text{HT}}$ admits a decomposition of the form in Eq.~\ref{eq_sigma_s}.
Such a structure arises, for example, in a perturbative treatment of the Holstein–Hubbard model, which describes interacting electrons coupled to phonons~\cite{PhysRevLett.122.026602}. 
There, the contribution of relatively high-energy excitations is identified as $\sigma_s$.
 In a different e-ph model, Ref.~\cite{werman2017non} similarly identifies a phonon-assisted channel which saturates at high temperature.
Despite differences, both studies illustrate how  Eq.~\ref{eq_sigma_s} emerges in e-ph models that lead to resistivity saturation.
Importantly, as we discuss below, this structure can also arise in regimes dominated by e-e scattering. 

Eq.~\ref{eq_prf} with $\sigma_{HT}$ obeying Eq.~\ref{eq_sigma_s} leads to  
\begin{equation}
    \big(\rho - \rho_{\text{R}}\big)^{-1} = \rho_i^{-1} + \rho_s^{-1},
\label{eq_prf2}
\end{equation}
where $\rho_i^{-1} = \sigma_i$ and $\rho_s^{-1} = \sigma_s$.  
Eq.~\ref{eq_prf2} resembles the so-called parallel-resistor formula, which is commonly used to  describe systems exhibiting resistivity saturation~\cite{PhysRevLett.38.782,hussey2003normal,hussey2004universality,PhysRevB.24.7404,doi:10.1126/science.1165015,PhysRevB.106.085141,SUNDQVIST2022110686}. 
The key distinction here is that, following Matthiessen’s rule [Eq.~\ref{eq_scarate}], we subtract from $\rho$ the contribution of mechanisms that do not produce saturation, namely $\rho_{\text{R}}$\footnote{This choice clarifies the limiting behaviors. In the usual parametrization one gets, as $T\!\to\!0$, $\rho\!\to\!\big(1/\rho_0+1/\rho_s\big)^{-1}$, which spuriously enforces a link between the residual resistivity and the high-temperature saturation limit. No such linkage arises in Eq.~\ref{eq_prf2}.}.
Furthermore, in Eq.~\ref{eq_prf2}, $\rho_i$ and $\rho_s$ are, by construction, 
associated with the same scattering mechanism: they account for different contributions to the corresponding conductivity—$\rho_i^{-1}$ captures the conventional, temperature-decreasing part, whereas $\rho_s^{-1}$ collects those that counteract this tendency at higher temperatures.

Having established how a parallel-resistor-like form can arise within this phenomenological framework, we now analyze Nb and LiFeAs using Eq.~\ref{eq_prf2} as our starting point. Its applicability to these materials is tested against the transport data; a microscopic derivation for their realistic electronic structures remains lacking. We do not exclude that alternative parametrizations could describe individual resistivity curves comparably well. As we show below, the value of the present framework lies in its physically motivated assumptions and in its ability to recover distinct transport behaviors in e-ph- and e-e-dominated systems while revealing a common transport regime across several iron-based superconductors. 

Applying Eq.~\ref{eq_prf2} to the data requires constraining $\rho_R$ and $\rho_s$. 
While plausible bounds for \(\rho_s\) can be inferred from the high-temperature data, estimating \(\rho_R\) is generally more delicate. Here we take \(\rho_R\) to be set by elastic impurity scattering, $\rho_R=\rho(T\to0)=\rho_0$~\footnote{Identifying $\rho_R$ with a constant $\rho_0$ is an additional approximation: a temperature-independent elastic scattering rate does not by itself ensure constant $\rho_R$ when the transport kernel $K(T,\Gamma)$ varies.}. This approximation is justified when no mechanism other than the one associated with \(\Gamma_{\mathrm{HT}}\) contributes appreciably to  \(\rho\). If, for example, e-e and e-ph scattering enter with comparable weight, the extracted \(\rho_i(T)\) would mix both and could not be cleanly assigned to either. Conversely, the ansatz is reasonable if the resulting \(\rho_i(T)\) displays canonical fingerprints of a single mechanism (e.g., \(T^2\) for Fermi-liquid e-e, or Bloch–Gr\"uneisen behavior for e-ph).

Appendix~\ref{app_fit} describes how $\rho_0$ and $\rho_s$ are constrained by the resistivity data: $\rho_0$ is primarily determined by the low-temperature behavior, whereas $\rho_s$ is mainly constrained by the curvature and flattening at higher temperatures. Because determining $\rho_s$ requires an ansatz for the as-yet-unknown $\rho_i(T)$, we use a flexible polynomial form at this stage and subsequently test whether our conclusions about $\rho_i$ remain robust under reasonable variations of $\rho_s$.
Once $\rho_0$ and $\rho_s$ are fixed, $\rho_i(T)$ is extracted pointwise from the measured resistivity by inverting Eq.~\ref{eq_prf2} at each temperature, without imposing any additional functional form on $\rho_i(T)$.
To characterize $\rho_i$, we use the logarithmic derivative $\alpha(T) = \frac{d\ln(\rho_i)}{d\ln(T)}$. 
When $\rho_i(T)$ exhibits a characteristic power law, $\alpha(T)$ approaches the corresponding exponent. 
The obtained parameters for all iron-based superconductors analyzed here are summarized in Table~\ref{table-fits} and the datasets used can be found in the repository~\cite{amigo}.

Fig.~\ref{fig:Nb_vs_Fe}(c) shows $\alpha(T)$ for Nb, evolving from $\sim1$ near room temperature to $\sim5$ at low $T$, while Fig.~\ref{fig:Nb_vs_Fe}(e) presents $\rho_i$, exhibiting a clear linear behavior above the Debye temperature, $\Theta_D \sim 270$\,K~\cite{PhysRev.181.1127}. 
These observations indicate that $\rho_i$ is well consistent with Bloch-Gr\"uneisen theory. 
Given that $\rho_i$ and $\rho_s$ are, by construction, associated with the same scattering mechanism, we conclude that e-ph scattering dominates the resistivity in Nb, including its tendency to saturation.

The main observation of this paper is shown in Fig.~\ref{fig:Nb_vs_Fe}(d), which reveals that for LiFeAs, remarkably, $\alpha(T)\sim2$ over nearly 300\,K. 
This exponent is consistent with e-e scattering in a Fermi liquid, which has been argued to dominate the resistivity of this compound in the low temperature limit~\cite{PhysRevB.85.060503}.
Fig.~\ref{fig:Nb_vs_Fe}(f) presents $\rho_i$ as a function of $T^2$, demonstrating that the $T^2$ behavior is controlled by a single prefactor over the entire temperature range. 
As demonstrated through several complementary tests in Appendix~\ref{app_fit}, the extracted $\rho_i$ remains dominated by a quadratic temperature dependence for all reasonable variations of $\rho_s$, showing that this result does not rely on a finely tuned value of the saturation scale.
Analysis of resistivity data from Refs.~\cite{PhysRevB.85.060503,PhysRevB.106.054508,zhang2019LiFeAs} yields the same phenomenology. 
Following the same rationale as in Nb, since $\rho_i$ and $\rho_s$ arise from the same scattering mechanism, we conclude that e-e scattering is also responsible for the sublinear-in-$T$ regime. 

\begin{figure}
    \centering
    \includegraphics[width=\columnwidth]{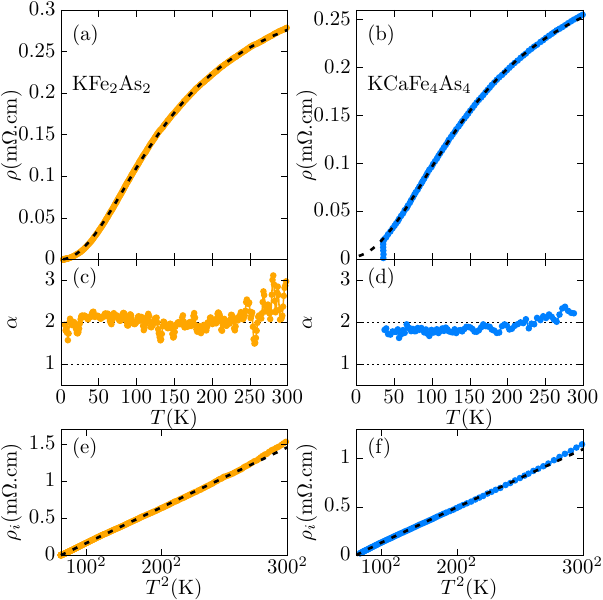}
    \caption{(a,b) Temperature dependence of the resistivity for KFe$_2$As$_2$ and KCaFe$_4$As$_4$, respectively. The data were digitized from Ref.~\cite{ doi:10.1143/JPSJ.79.124713, PhysRevB.96.140501}. The dotted line is a fit of Eq.~\ref{eq_prf2}. (c,d) Logarithmic derivative of $\rho_i$ as a function of temperature for both compounds. (e,f) Extracted $\rho_i$ as a function of $T^2$. The dotted line is a linear fit of the data.}
    \label{fig:other_systems}
\end{figure}

\section{Other systems}
\label{sec:other}
Iron-based superconductors comprise different families of compounds, so it is natural to ask whether other members exhibit a phenomenology similar to that of LiFeAs.  
Indeed, there are cases that display resistivity saturation but also show qualitative differences, for example in the values of the residual resistivity or in the detailed temperature dependence of $\rho$~\cite{bhoi2008resistivity,PhysRevB.83.212506}.  
Key features of LiFeAs are that it does not develop long-range magnetic order and that it remains in the tetragonal phase down to the superconducting transition.  
This motivates us to examine first other systems within the same class.

Figs.~\ref{fig:other_systems}(a),(c),(e) present data for KFe$_2$As$_2$ from Ref.~\cite{doi:10.1143/JPSJ.79.124713}, while
Figs.~\ref{fig:other_systems}(b),(d),(f) show the corresponding results for KCaFe$_4$As$_4$~\cite{PhysRevB.96.140501}. As in LiFeAs, the resistivity is well described within the proposed framework, with $\rho_i$ dominated by a quadratic term. The characteristic exponent $\alpha(T)$ rises slightly above $2$ for $T>250$\,K. 

We examine the sensitivity to a temperature-dependent saturating conductivity in Appendix~\ref{app_sigmas}. Allowing this additional freedom preserves the nearly quadratic behavior of $\rho_i$, although the extent to which this additional freedom is used by the fit is material dependent. For LiFeAs and KFe$_2$As$_2$, the fitted $\sigma_s$ varies only by $\sim3\%$ and $\sim1\%$ across the respective fitting windows. For KCaFe$_4$As$_4$, the variation reaches $\sim12\%$, and the upward deviation of $\alpha(T)$ above $250$\,K is removed by the modified extraction. We note, however, that this change does not, by itself, identify a temperature-dependent saturating conductivity as the physical origin of the deviation. Temperature-dependent spectral-weight transfers or a temperature-dependent quasiparticle weight---both encoded in the transport kernel $K(T,\Gamma)$---may also become relevant in this regime. In addition, as $\rho-\rho_0$ approaches $\rho_s$, the values of $\rho_i$ extracted using Eq.~\ref{eq_prf2} become particularly sensitive to small changes in $\rho_s$. The resulting high-temperature scatter in $\alpha(T)$ is especially evident for KFe$_2$As$_2$. 
Regardless of these considerations, we emphasize that there is no evidence of the pronounced crossover in $\alpha(T)$ expected if the e-ph contribution were dominant, as exemplified by Nb.

\begin{figure}
    \centering
    \includegraphics[width=\columnwidth]{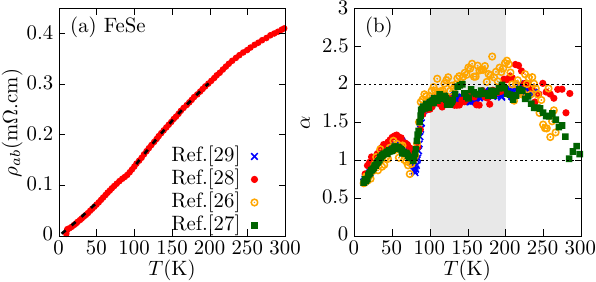}
	\caption{(a) Temperature dependence of the resistivity of FeSe for the $ab$-plane from data of Ref. \cite{kasahara2014field}. The dotted lines are fits of Eq.~\ref{eq_prf2}.
	(b) Logarithmic derivative of $\rho_i$ as a function of the temperature, based on data from Refs. \cite{PhysRevB.110.224508, tanatar2016origin, kasahara2014field, watson2015emergence}.}
    \label{fig:FeSe}
\end{figure}

In this context, it is of interest to analyze FeSe, which is well known for its nematic transition at $T_s = 90$\,K~\cite{watson2015emergence,shimojima2014lifting,fernandes2014drives, fedorov2016effect,massat2016charge,baek2015orbital,PhysRevLett.114.027001,PhysRevX.6.021032,PhysRevLett.116.227001,wang2015strong,wang2015nematicity, Glasbrenner2015effect,wang2016magnetic,PhysRevB.94.155138, PhysRevX.9.041049}, where the crystal structure changes from tetragonal to orthorhombic.  
In the nematic phase, the resistivity is approximately linear in $T$, 
and scattering from spin fluctuations is significant~\cite{PhysRevResearch.2.013309,PhysRevB.110.224508}. 
Above $T_s$, it shows a tendency toward saturation [Fig.~\ref{fig:FeSe}(a)]~\cite{0953-2048-28-10-105009,abdel2016impurity}  
 and the nature of the dominant scattering mechanism has remained unsettled.  
Fig.~\ref{fig:FeSe}(b) shows the results of applying Eq.~\ref{eq_prf2} to datasets from different groups: in all cases $\alpha(T)$ changes abruptly from values close to 1 in the nematic phase to 2 in the tetragonal phase.  
This value remains roughly constant over nearly 100\,K above $T_s$ (depending on the considered dataset). Within this range, FeSe behaves similarly to the previously discussed compounds.

A few remarks are in order regarding the nematic phase itself, where the exponent $\alpha(T)$ is not pinned to unity but exhibits a systematic temperature dependence [Fig.~\ref{fig:FeSe}(b)]. 
First, for sufficiently low temperatures the decomposition introduced by Eq.~\ref{eq_prf2} plays essentially no role: there $\rho-\rho_R\ll\rho_s$, so the parallel combination reduces to $\rho_i\approx \rho-\rho_R$ and $\alpha(T)$ simply reflects the temperature dependence of the ($\rho_R$-subtracted) resistivity.
Second, the situation is different, and considerably more  intricate, for temperatures approaching $T_s$. In that regime, transport is expected to become more complicated owing to the possible relevance of scattering from strong nematic fluctuations, which are coupled to the lattice and
hybridize with phonons into a mixed electronic--lattice  mode~\cite{cano2010interplay,karahasanovic2016elastic,PhysRevLett.118.227601}. 
Accordingly, we cannot exclude that the sharp variation of $\alpha(T)$ near
$T_s$ partly reflects a breakdown of some of the assumptions underlying
Eq.~\ref{eq_prf2}, in particular the simple separation between el-el and el-ph
contributions. Even in that case, however, the loss of validity of the
decomposition would itself signal the emergence of a distinct transport regime
associated with nematic order.

Above $T_s$, the influence of nematic fluctuations is expected to diminish progressively with increasing temperature. Our central finding---that FeSe
rapidly recovers an $\alpha\approx2$ behavior analogous to that of the
non-nematic compounds---concerns this extended tetragonal regime, where the
assumptions underlying the decomposition are expected to be better controlled.

  A word of caution about the generality of this $\alpha\approx2$ regime among tetragonal systems is also in order. Suppressing the structural transition does not, by itself, guarantee   LiFeAs-like transport over an extended temperature  range. A pertinent example is FeSe$_{0.75}$S$_{0.25}$, in which there is no structural transition and quadratic normal-state resistivity is recovered below approximately $9\,\mathrm{K}$ when superconductivity is suppressed by a magnetic field~\cite{PhysRevResearch.2.013309}. In the temperature range 10--140\,K, however, a linear fit to its resistivity derivative yields an appreciable positive intercept upon extrapolation to \(T=0\), as in FeSe and in contrast to LiFeAs, whose fitted intercept is consistent with zero (Appendix~\ref{app_fit}).  Persistent nematic fluctuations could contribute to the contrasting  transport of the sulfur-substituted compound despite the absence of long-range nematic order, although other  material-dependent effects cannot be excluded.

At the high-temperature end of FeSe, the description faces a
different limitation. For $T\gtrsim220\,$K, Eq.~\ref{eq_prf2} with a
temperature-independent $\rho_s$ no longer provides a satisfactory description
of the data. 
The more pronounced variability among datasets at high temperatures, together
with the sample-quality dependence reported in Ref.~\cite{PhysRevB.94.024526},
motivates considering extrinsic effects such as impurities or internal strain. 
The high-temperature deviations may also have an intrinsic contribution, as ARPES studies have reported substantial temperature-driven evolution of the electronic structure~\cite{kushnirenko2017anomalous,rhodes2017strongly}. The present analysis does not distinguish between these possibilities.

Despite the distinct complications affecting the analysis below $T_s$ and above $\sim220\,$K, it is remarkable that in the intervening range ($T_s\lesssim T\lesssim220\,$K) FeSe exhibits a saturating-$T^2$ resistivity, with $\alpha\approx2$, strikingly similar to that of the non-nematic compounds analyzed above. This common behavior—together with its abrupt modification across the nematic transition in FeSe—points to a generic transport regime. It is therefore natural to ask whether $\rho_s$ is an intrinsic property of a material, analogous to the Fermi-liquid prefactor ($A_{2}$) that correlates with thermodynamic quantities (e.g., via the Kadowaki--Woods relation).
Here, following Ref.~\cite{wang2020t,https://doi.org/10.1002/andp.202100588}, we use an extension of this approach, convenient for semimetallic systems with small Fermi energies, where instead of the electronic specific heat, the scaling involves the Fermi energy ($E_F$). Fig.~\ref{fig:A2}(a) summarizes the values of \(A_{2}\) for several compounds from different groups, as obtained in Appendix~\ref{app_fit}; the magnitudes follow the trend expected for a Fermi liquid~\cite{https://doi.org/10.1002/andp.202100588}. The pocket-to-pocket spread in $E_F$, indicated by the error bars, is small compared with the overall range of Fermi energies in Fig.~\ref{fig:A2}(a) and does not obscure the qualitative trend.

Fig.~\ref{fig:A2}(b) displays \(A_{2}\) as a function of \(\rho_{s}\). Within each compound, the data collapse onto an approximately straight line passing through the origin,
$A_{2}=\nu \rho_{s}$,
with a material-dependent slope \(\nu\). If $A_2$ is
intrinsic, the scatter of reported $A_2$ values for a given
compound most plausibly reflects uncertainties in absolute resistivity determination. The observed linear proportionality then suggests that the corresponding spread
in $\rho_s$ has the same origin.
Importantly, the material-dependent slope $\nu$ is independent of the geometric factor involved in the resistivity determination.
Namely, a residual geometric correction factor $g_j$ may rescale both parameters for sample $j$:
\begin{equation}
A_{2,j}^{\rm meas}=g_jA_{2,j}^{\rm true},
\qquad
\rho_{s,j}^{\rm meas}=g_j\rho_{s,j}^{\rm true}.
\end{equation}
The observed proportionality
$A_{2,j}^{\rm meas}=\nu\rho_{s,j}^{\rm meas} $  implies
$A_{2,j}^{\rm true}=\nu\rho_{s,j}^{\rm true} $, independently of $g_j$. 
Therefore, under the assumption that $A_2^{\rm true}$ is sample-independent within
each compound, the observed linear relation supports an intrinsic, sample-independent $\rho_{s} ^{ \rm true}$.

Equivalently, the slope defines a characteristic temperature $T_{\nu}=\nu^{-1/2}=\sqrt{\rho_s/A_2}$, at which, in a given material, $A_2T_\nu^2=\rho_s$; above this scale, the saturating channel increasingly dominates the total conductivity. Because $T_{\nu}$ depends only on the ratio of $\rho_s$ and $A_2$, it is unaffected by $g_j$ and can therefore be compared directly across samples and groups without requiring knowledge of the sample dimensions. The central values in Table~\ref{table-fits} form four nonoverlapping intervals: 138--149\,K (KFe$_2$As$_2$), 164--178\,K (KCaFe$_4$As$_4$), 191--201\,K (FeSe), and 209--229\,K (LiFeAs). The spread within each compound (10--20\,K) is typically smaller than the separation between the central values for different compounds.

\begin{figure}
    \centering
    \includegraphics[width=1\columnwidth]{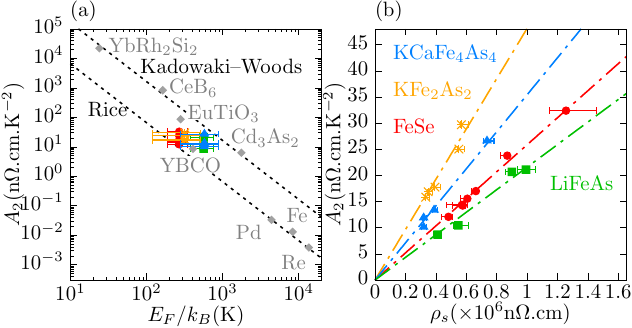}
	\caption{(a) $T^2$ coefficient vs Fermi energy scaling. The gray points are reproduced from Ref.\cite{wang2020t}. $A_2$ values for the iron based superconductors are obtained from $\rho_i$ while  $E_F$ is obtained from Ref. \cite{PhysRevB.97.134501, terashima2014anomalous, doi:10.1143/JPSJ.79.053702, PhysRevLett.108.047002} %
    as explained in Appendix~\ref{app-EF}.
	Dashed lines correspond to the Kadowaki-Woods and Rice limits. (b) $A_2$ as a function of $\rho_s$. Values are obtained using Eq. \ref{eq_prf2} with digitized data from the references \cite{PhysRevB.110.224508, tanatar2016origin, kasahara2014field, watson2015emergence,zhou2021FeSe} (FeSe),  \cite{hardy2013evidence, nakajima2014strong, 10.1143/JPSJ.78.063702, doi:10.1143/JPSJ.79.124713,fang2015KFe2As2} (KFe$_2$As$_2$), \cite{morozov2010single, PhysRevB.85.060503, PhysRevB.106.054508,zhang2019LiFeAs} (LiFeAs) and \cite{PhysRevB.94.064501, PhysRevB.96.140501,singh2018CaKFe4As4,pyon2019CaKFe4As4} (KCaFe$_4$As$_4$).}    \label{fig:A2}
\end{figure}

\section{Discussion}
\label{sec:discussion}

We have presented a phenomenological analysis of the resistivity of various iron-based superconductors. Our approach consists of two main steps. The first, based on Matthiessen's rule, separates the scattering mechanism responsible for the sublinear-in-$T$ resistivity. The second, heuristically, decomposes the conductivity associated with this mechanism into a conventional metallic component, $\sigma_i$, and a temperature-independent term, $\sigma_s$ [Eq.~\ref{eq_sigma_s}].  This leads to an expression analogous to the widely used parallel-resistor formula, but with the conceptual advantage of explicitly identifying the contribution from the mechanism causing saturation. 
By the same rationale that establishes e-ph scattering as dominant in Nb, we associate the sublinear-in-$T$ regime in LiFeAs with e-e scattering.

A natural question is whether e-ph scattering plays a more prominent role in the $T^2$ resistivity. While this trend is commonly attributed to e-e scattering, recent observations 
in SrVO$_3$ and other oxides have reignited debates about the relevance of e-ph scattering in correlated metals~\cite{lee2024fermi,PhysRevLett.133.186501,PhysRevMaterials.7.093801, PhysRevB.110.L220504,mirjolet2021electron}. In addition, systems with cylindrical Fermi surfaces may have an intermediate temperature regime where e-ph leads to a $T^2$ behavior~\cite{PhysRevB.18.1849}.

Recently, Ref.~\cite{PhysRevLett.133.186501} reports that density-functional plus dynamical mean-field theory (DFT+DMFT) calculations for SrVO$_3$ yield a resistivity scaling as $T^{1.9}$ between 100\,K and 200\,K, despite dominant e-ph scattering.
A linear-in-$T$ e-ph resistivity sets in only above 300\,K, and this rather delayed onset of the $T$-linear regime is attributed to high-frequency optical phonon modes, which are weakly populated at lower temperature.
In LiFeAs, however, DFT calculations indicate that the phonon bandwidth is approximately half that of SrVO$_3$~\cite{https://doi.org/10.1155/2010/804343}, 
suggesting that the highest-in-energy optical phonons become populated at significantly lower temperatures.

In addition, approaching the low temperature limit, backscattering by acoustic phonons is expected to become increasingly ineffective below the Bloch-Gr\"uneisen temperature scale 
$k_B T_{BG} = 2\hbar v_s k_F$, where $k_F$ is the Fermi momentum and $v_s$ the sound velocity.
For LiFeAs, a lower bound yields $T_{BG} \sim 120$\,K (see Appendix~\ref{app_estimates}).
The persistent $T^2$ dependence well below this scale, with no sign of a crossover to higher powers, suggests that e-ph scattering is subleading.
We note that a predominance of e-e over e-ph scattering at elevated temperatures has also been reported in other Hund's metals~\cite{PhysRevMaterials.7.093801}  while recent Hubbard–Holstein DMFT calculations show that electronic correlations can strongly suppress the effective e-ph coupling while leaving the low-energy \(T^2\) electronic scattering largely unaffected~\cite{3qk3-py1n}. 

One scenario in which a sublinear-in-$T$ regime originates from e-e scattering is the coherence–incoherence crossover discussed in different correlated systems~\cite{haule2009coherence,PhysRevLett.106.096401}. 
In KFe$_2$As$_2$, evidence for such a crossover comes from the evolution of the magnetic susceptibility from a low-temperature Pauli-like response to a Curie--Weiss-like behavior at elevated temperatures, accompanied by a pronounced maximum in the electronic thermal expansion~\cite{hardy2013evidence}. Consistently, NMR measurements reveal a crossover from coherent quasiparticle behavior to a local-moment-like regime~\cite{PhysRevB.97.064509}. Optical conductivity measurements provide further evidence:
a coherent response emerges below approximately 155~K,
while its scattering rate follows a $T^2$ dependence only
below approximately 75~K~\cite{PhysRevB.96.201108}. A related crossover has also been proposed for LiFeAs based on a change in the curvature of its resistivity. Its evolution across the Li(Fe,Co)As series has been examined alongside doping-dependent quasiparticle mass renormalizations measured by ARPES and DMFT calculations~\cite{PhysRevB.103.054503}.  

Theoretical studies in Hund's metals show that, as temperature increases, coherent Fermi-liquid quasiparticles gradually lose spectral weight and their lifetime shortens, yet remnants of these excitations persist above the Fermi-liquid regime. 
These short-lived but still dispersive carriers—termed \emph{resilient quasiparticles}~\cite{PhysRevLett.106.096401,PhysRevLett.110.086401}—govern transport in an extended intermediate regime in which the self-energy no longer follows Fermi-liquid scaling.

In this picture, the deviation from the $T^2$ resistivity is associated with the coherence-incoherence crossover.
For LiFeAs, one can estimate a coherence scale $T^*$ from the measured low-temperature resistivity by requiring $\Gamma(T^*)/k_{\mathrm{B}}T^* \!\sim\! 1$. 
Based on a model of the pockets measured in quantum-oscillation experiments~\cite{PhysRevLett.108.047002}, one estimates $T^*\sim20$--$35$\,K (see Appendix~\ref{app_estimates}).
Significant deviations from the $T^2$ law become apparent only at higher temperatures, around $40$--$45$\,K.
This is reminiscent of the “hidden Fermi-liquid’’ picture of Ref.~\cite{PhysRevLett.111.036401} in which, before reaching saturation, the scattering rate $\Gamma$ can continue to grow as $T^2$ above the strict Fermi-liquid regime.

A simple parametrization for the saturation of the scattering rate is $\Gamma(T)=\hbar/(\tau_{\mathrm{FL}}+\tau_s)$,  where $\tau_{\mathrm{FL}}\!\propto\!1/T^2$ and  $\tau_s$ represents a short, weakly temperature-dependent lifetime.
As detailed in Appendix~\ref{app_toymodel}, this form naturally yields a decomposition of the conductivity as in Eq.~\ref{eq_sigma_s}, with $\sigma_i \propto K(T,\Gamma)  \tau_{\mathrm{FL}}$ and $\sigma_s \propto  K(T,\Gamma) \tau_s$. 
Notably, the consideration of an energy-independent scattering rate, here adopted for simplicity, can be too crude in real materials.
More generally, by analogy with treatments in the e-ph literature, where a contribution to the conductivity expressed as an integral over excitation energies is identified as the source of saturation~\cite{PhysRevLett.122.026602}, one may view $\sigma_s$ as an effective contribution to the conductivity associated with short-lived electronic excitations in the coherence–incoherence crossover.

Even if $\Gamma$ tends to saturate at high temperatures, the resistivity need not do so~\cite{PhysRevLett.106.096401,PhysRevLett.111.036401}. 
The temperature dependence of the quasiparticle residue and the redistribution of spectral weight can naturally modify the transport response. 
The observation that a decomposition of the form Eq.~\ref{eq_sigma_s}, with a seemingly constant $\sigma_s$ term, accurately describes the conductivity over a remarkably wide temperature range in several iron-based superconductors provides a clear motivation for future theoretical studies.

The case of FeSe offers a particularly interesting perspective.
Similar to LiFeAs, KFe$_2$As$_2$, and KCaFe$_4$As$_4$, our analysis indicates
that transport in the tetragonal phase is dominated by e-e scattering and is
consistent with the coherence--incoherence crossover. Although the nematic
transition is widely regarded as being of electronic origin, the tetragonal
state immediately above $T_s$ is often treated as a coherent metal. Our
results instead suggest that the nematic transition develops out of an already
partially incoherent electronic state.

\section{Conclusions}
\label{sec:conclusions}

In summary, we have identified a saturating-$T^2$ resistivity regime common to the tetragonal phase of several iron-based superconductors.
The saturating tendency is governed by a contribution to the conductivity which, when taken as temperature independent, reproduces the experimental data over a remarkably broad temperature range.
We interpret this contribution as arising from short-lived electronic excitations, consistent with the coherence–incoherence crossover discussed in Hund’s metals.

While such crossovers have been discussed previously,
our analysis provides a complementary transport
perspective. By combining Matthiessen's rule with a
phenomenological additive-conductivity decomposition,
we uncover an extended, nearly quadratic contribution
that is obscured in the measured resistivity. The
comparison with Nb supports an electronic, rather than
predominantly phononic, origin of the saturating tendency
in the iron-based superconductors considered here.
Furthermore, the observed proportionality between
the quadratic coefficient $A_2$ and the saturation
parameter $\rho_s$ suggests that $\rho_s$ is an intrinsic
material parameter, provided that $A_2$ is itself
sample-independent.

The similar transport behavior of FeSe above the nematic transition and the other compounds suggests a common incoherent high-temperature regime, with the nematic symmetry breaking in FeSe and the recovery of a coherent Fermi-liquid regime in other systems representing two distinct routes by which this incoherent state is resolved at low temperature.

\section{Acknowledgments.}

We thank Gladys Nieva and Pablo S. Cornaglia for useful discussions. 

\appendix

\section{Fitting procedure}
\label{app_fit}

The analysis of Sec.~\ref{sec:saturation} requires fixing only two quantities from the data before $\rho_i(T)$ and its logarithmic derivative $\alpha(T)$ can be obtained: the non-saturating contribution $\rho_R$, which we identify with the residual resistivity $\rho_0$, and the saturating scale $\rho_s$. Once these two numbers are fixed, Eq.~\ref{eq_prf2} is inverted pointwise,
\begin{equation}
\rho_i(T)=\bigl[\,(\rho(T)-\rho_0)^{-1}-\rho_s^{-1}\bigr]^{-1},
\label{eq_rhoi_extract}
\end{equation}
with no additional functional form imposed on $\rho_i(T)$.
The logarithmic derivative $\alpha(T)=\mathrm{d}\ln\rho_i/\mathrm{d}\ln T$ is then obtained by numerical differentiation. The purpose of this appendix is to describe how $\rho_0$ and $\rho_s$ are constrained.

\emph{Low-temperature limit.} In this limit, $\rho\ll\rho_s$, so that $\rho(T)\simeq\rho_0+\rho_i(T)$ and $\rho_0=\rho(T\!\to\!0)$. We therefore determine $\rho_0$ by extrapolating a polynomial fit of the low-temperature resistivity $\rho(T)\simeq\rho_0+\sum_i B_iT^i$.

\begin{figure*}
    \centering
    \includegraphics[width=\textwidth]{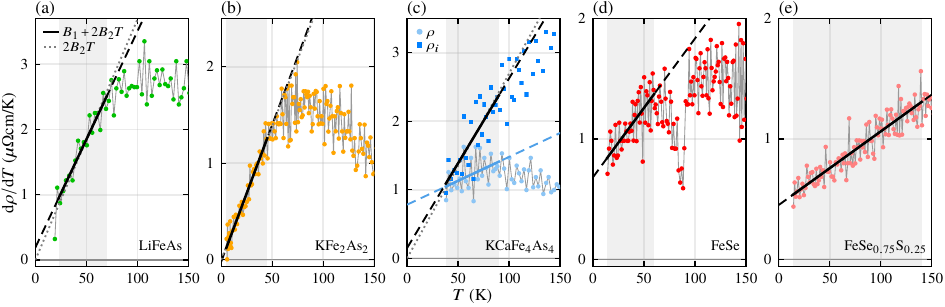}
    \caption{$\mathrm{d}\rho/\mathrm{d}T$ versus temperature for (a) LiFeAs, (b) KFe$_2$As$_2$, (c) KCaFe$_4$As$_4$, (d) FeSe and (e) FeSe$_{0.75}$S$_{0.25}$. In every panel the straight line is a fit $\mathrm{d}\rho/\mathrm{d}T=B_1+2B_2T$ over the shaded low-temperature window, drawn solid where it is fitted and dashed outside it. In (a)--(c) the dotted line is the same fit constrained to pass through the origin ($B_1\equiv0$). In (c), the measured derivative (light circles) exhibits negative curvature throughout the accessible normal-state range above $T_c\simeq35$\,K. For this reason, $\rho_0$ and $\rho_s$ were determined self-consistently by including the saturating contribution in the low-temperature fit. In (d) the drop and recovery of $\mathrm{d}\rho/\mathrm{d}T$ across the nematic transition of FeSe at $T_s\simeq90$\,K is apparent.}
    \label{fig:app_extraction}
\end{figure*}

Whether a quadratic or a linear dependence dominates this expansion can be seen directly in the temperature derivative of the raw resistivity. For LiFeAs, Fig.~\ref{fig:app_extraction}(a) shows that $\mathrm{d}\rho/\mathrm{d}T$ is linear in $T$ up to $\sim60$\,K and extrapolates through the origin. A free fit, $\mathrm{d}\rho/\mathrm{d}T=B_1+2B_2T$, gives $B_1=(1.9\pm1.4)\times10^{-4}$\,m$\Omega$\,cm\,K$^{-1}$, consistent with zero and hence with a negligible linear-in-$T$ resistivity contribution. The same fit applied to KFe$_2$As$_2$ returns intercepts within two standard deviations of zero for all the considered datasets [Fig.~\ref{fig:app_extraction}(b)]. In FeSe, the orthorhombic phase contains a sizable linear-in-$T$ contribution [Fig.~\ref{fig:app_extraction}(d)].  For the Nb benchmark, the relevant low-temperature terms are instead $T^3$ and $T^5$, as suggested by Ref.~\cite{PhysRev.181.1127}.

KCaFe$_4$As$_4$, with its comparatively higher critical temperature $T_c\simeq35$\,K, is a special case~\cite{PhysRevB.97.140508}. In this system a contribution that bends the resistivity downward from a pure $T^2$ law is already active immediately above $T_c$ [Fig.~\ref{fig:app_extraction}(c), light circles], so the approximation $\rho\ll\rho_s$ underlying the low-$T$ fit does not hold over a sufficiently broad temperature range. We therefore determined $\rho_0$ and $\rho_s$ self-consistently for this compound, keeping the same quadratic form but including the saturating channel in the low-temperature fit [Fig.~\ref{fig:app_extraction}(c), full squares]. Allowing instead a negative cubic term to describe the downwards bending in a bare low-$T$ fit describes the data equally well and changes the resulting $\alpha(T)$ by less than $0.1$, so the analysis does not depend on which of the two is adopted.

The comparison with FeSe$_{0.75}$S$_{0.25}$ is particularly  informative because sulfur substitution suppresses the   nematic transition. Despite this  suppression, its resistivity derivative retains a   qualitative similarity to that of parent FeSe: linear  fits yield appreciable positive intercepts over the   examined temperature windows [Fig.~5(d,e)], whereas  the LiFeAs intercept is consistent with zero  [Fig.~5(a)]. In this respect, the sulfur-substituted   compound remains closer to FeSe than to LiFeAs.   Thus, suppressing long-range nematic order does not,  by itself, produce LiFeAs-like transport over an extended   temperature range. This comparison concerns the examined
finite-temperature windows and is compatible with the   quadratic normal-state resistivity reported below   approximately $9\,\mathrm{K}$ when superconductivity is suppressed by a magnetic field for   FeSe$_{0.75}$S$_{0.25}$~\cite{PhysRevResearch.2.013309}.

\emph{Determination of $\rho_s$.} The saturating scale is set by the high-temperature data. With $\rho_0$ fixed as above, we fit the full expression $\rho (T)=\rho_0+(\rho_i(T)^{-1}+\rho_s^{-1})^{-1}$ to $\rho(T)$, taking $\rho_i(T)$ in the \emph{general} form $\rho_i(T)=A_1T+A_2T^2+A_3T^3$, with the $A_i$ and $\rho_s$ all free and non-negative. 
The least-squares optimum based on this fitting ansatz defines the single value of $\rho_s$ (one per dataset) used throughout the main text; for the LiFeAs data it gives $\rho_s\approx0.9$\,m$\Omega$\,cm. The resulting quadratic coefficients $A_2$ are typically close to the corresponding $B_2$ values obtained from the low-temperature fits.

Over the temperature window used to fit FeSe, 95–200\,K, the temperature varies by only a factor of approximately 2.
Consequently, the three powers of $T$ are not sufficiently independent as fitting functions: a combination of $T$ and $T^3$ can mimic $T^2$ behavior over such a limited range. For the datasets in which this degeneracy becomes apparent, we therefore fix $A_1/A_2=c_{\rm ref}=22.95$\,K, the value obtained from Ref.~\cite{watson2015emergence}, for which this ratio is best determined ($\pm2$\,K over 95--250\,K). Whether this ratio is fixed or fitted freely, the logarithmic derivative $\alpha(T)$ remains close to 2.

In Nb, for the high-temperature region  1000--2000\,K, we used Eq.~\ref{eq_prf2}, with $\rho_i= A_1T$ and $\rho_R=\rho_0$.
The quadratic component from the e-e interaction should be the same as that estimated at low temperature and remains negligible even at $T\sim2000$\,K, so that including only a linear term in $\rho_i$ is a good approximation.  
The resulting values are $\rho_0=(0.020\pm0.003)$\,$\mu\Omega$\,cm and $\rho_s=(0.279\pm0.006)$\,m$\Omega$\,cm, based on the data of Refs.~\cite{PhysRev.181.1127,abraham1972resistivite}.

\emph{Correlation between $\alpha$ and $\rho_s$.}
The procedure defined above to determine $\rho_s$ does not privilege a particular term in $\rho_i$: $T$, $T^2$ and $T^3$ are free to compete. An important question is whether the extracted exponent $\alpha(T)$ depends sensitively on the precise value of $\rho_s$.
We show next that this is not the case and that, in fact,  the quadratic dominance of $\rho_i$ is a robust feature of the data. We support this with two complementary tests: a comparison of fitting forms for LiFeAs and a sensitivity analysis for one representative dataset of each of LiFeAs, KFe$_2$As$_2$, KCaFe$_4$As$_4$, and FeSe.

First, fitting Eq.~\ref{eq_prf2} for LiFeAs \cite{morozov2010single} under a range of assumptions for $\rho_i$ shows that every model containing a $T^2$ term describes the data well over the entire range and returns a consistent $\rho_s\approx0.9$\,m$\Omega$\,cm, whereas a model that deliberately omits the $T^2$ term fails: the fit is forced to $\rho_s=0.77$\,m$\Omega$\,cm, barely above the maximum measured resistivity ($\approx0.64$\,m$\Omega$\,cm at $313$\,K), which imposes near-complete saturation already at the highest measured temperatures and pushes the fit systematically below the data. A single-term fit with free exponent, $\rho_i=A_nT^{n}$, gives $n=1.97\pm0.01$. Moreover, if $\rho_s$ is fit using $\rho_i=A_1T+A_2T^2+A_3T^3$, the $T^2$ term supplies $\sim99.8\%$ of $\rho_i$ at $200$\,K, with the $T^3$ coefficient indistinguishable from zero.

\begin{figure*}
    \centering
    \includegraphics[width=18cm]{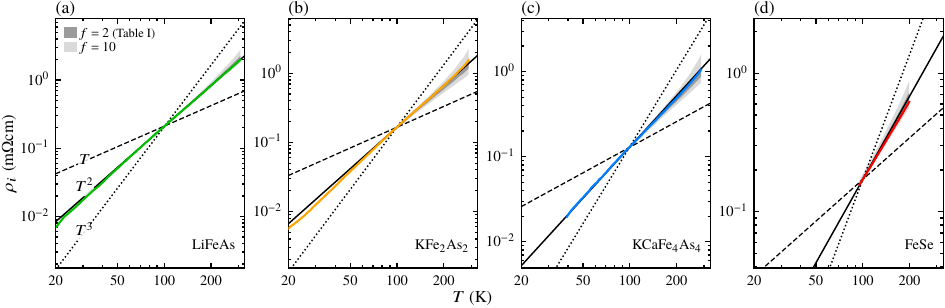}
    \caption{Extracted $\rho_i$ [Eq.~\ref{eq_rhoi_extract}] for one representative dataset of each compound: (a) LiFeAs~\cite{morozov2010single}, (b) KFe$_2$As$_2$~\cite{doi:10.1143/JPSJ.79.124713}, (c) KCaFe$_4$As$_4$~\cite{PhysRevB.96.140501} and (d) FeSe~\cite{watson2015emergence}. Coloured curves use the $\rho_0$ and $\rho_s$ of Table~\ref{table-fits}; the black lines are $T$, $T^2$ and $T^3$ references. The shaded bands show how the extracted $\rho_i$ changes as $\rho_s$ is varied. Each band is defined by the factor $f$, which sets the allowed increase in the unexplained variance of the polynomial fit relative to its best value: up to twice that value for the inner band ($f=2$), and ten times for the outer band ($f=10$). The half-width of the $\rho_s$ range used for the inner band is quoted as its uncertainty in Table~\ref{table-fits}. }
    \label{fig:app_rhoi}
\end{figure*}

Second, the quadratic dominance persists also if $\rho_s$ is allowed to vary from its optimal fitted value. Fig.~\ref{fig:app_rhoi} shows the extraction of Eq.~\ref{eq_rhoi_extract} for different values of $\rho_s$, with $\rho_0$ fixed from the low-temperature limit. For each fixed $\rho_s$, the extracted $\rho_i$ is fitted to $A_1T+A_2T^2+A_3T^3$, with all three amplitudes free and non-negative, and its unexplained variance, $1-R^2$, is evaluated. We define a shaded band as the envelope of the $\rho_i(T)$ curves for which this variance remains within a factor $f$ of its minimum value.
The optimum $\rho_s$ is determined by fitting the measured resistivity, whereas the bands illustrate the sensitivity of the extracted $\rho_i$ to changes in $\rho_s$, using the quality of its polynomial description as the selection criterion.
Although the $T^2$ reference line lies close to the curve obtained with the optimal $\rho_s$, the bands remain close to quadratic behavior, showing that the quadratic predominance is not an outcome of fine-tuning $\rho_s$.

Taken together, all these observations show that the functional form of $\rho_i$ assumed while constraining $\rho_s$ affects only weakly the extracted $\alpha(T)$.

\begin{table*}[t]
\begin{center}
\caption{Parameters obtained from the fit of Eq.~\ref{eq_prf2} to each dataset, with $\rho_i=A_1T+A_2T^2+A_3T^3$ and $\rho_0$ determined as described in the text, including the self-consistent treatment of KCaFe$_4$As$_4$. %
$T_{\nu}=\sqrt{\rho_s/A_2}$. Units are $\times 10^{-3}$m$\Omega$.cm for $\rho_0$, m$\Omega$.cm for $\rho_s$, and  $\times 10^{-6}$m$\Omega$.cm.K$^{-n}$ for $A_n$. Uncertainties on $\rho_0$ are the spread over the choice of a low-temperature window; those on $\rho_s$ are the half-width of the range over which the polynomial description of the extracted $\rho_i$ remains of comparable quality to that obtained at the optimum; those on $A_2$ combine the fit error with the drift of $A_2$ across that range. Uncertainties on $A_1$ and $A_3$ are those of the fit.  The last column gives the temperature range of the fit. S1-S3 in FeSe are labeled as in Ref.~\cite{zhou2021FeSe} }\label{table-fits}
\begin{ruledtabular}
\begin{tabular}{l c c c c c c c}
Material & $\rho_0$ & $\rho_s$ & $A_1$ & $A_2$ & $A_3$ & $T_{\nu}$(K) & $T$ range (K) \\
\hline
\multirow{4}{*}{LiFeAs} & $20\pm2$ & $0.90\pm0.03$ & $7\pm20$ & $21\pm1$ & $0.000\pm0.003$ & $209\pm3$ & 18--313 \cite{morozov2010single} \\
 & $11\pm2$ & $0.99\pm0.06$ & $60\pm30$ & $21\pm1$ & $0.000\pm0.005$ & $217\pm10$ & 18--290 \cite{PhysRevB.85.060503} \\
 & $5.0\pm0.4$ & $0.55\pm0.05$ & $23\pm10$ & $10.4\pm0.9$ & $0.000\pm0.002$ & $229\pm20$ & 18--290 \cite{PhysRevB.106.054508} \\
 & $5.1\pm0.7$ & $0.41\pm0.02$ & $26\pm6$ & $8.7\pm0.5$ & $0.000\pm0.001$ & $217\pm6$ & 18--290 \cite{zhang2019LiFeAs} \\
\hline
\multirow{5}{*}{KFe$_2$As$_2$} & $0\pm1$ & $0.35\pm0.02$ & $10\pm30$ & $17\pm2$ & $0.000\pm0.006$ & $144\pm10$ & 5--290 \cite{nakajima2014strong} \\
 & $0.5\pm0.2$ & $0.55\pm0.02$ & $0\pm30$ & $25\pm2$ & $0.005\pm0.007$ & $148\pm7$ & 5--290 \cite{hardy2013evidence} \\
 & $6\pm2$ & $0.57\pm0.03$ & $0\pm20$ & $30\pm2$ & $0.004\pm0.005$ & $138\pm8$ & 5--290 \cite{10.1143/JPSJ.78.063702} \\
 & $0.3\pm0.6$ & $0.33\pm0.01$ & $0\pm20$ & $15.8\pm0.9$ & $0.005\pm0.003$ & $145\pm6$ & 5--290 \cite{doi:10.1143/JPSJ.79.124713} \\
 & $0.3\pm0.5$ & $0.39\pm0.02$ & $0\pm20$ & $18\pm2$ & $0.000\pm0.004$ & $149\pm10$ & 5--290 \cite{fang2015KFe2As2} \\
\hline
\multirow{4}{*}{KCaFe$_4$As$_4$} & $4.2\pm0.1$ & $0.39\pm0.02$ & $37\pm20$ & $13.4\pm0.9$ & $0.000\pm0.004$ & $170\pm9$ & 38--300 \cite{PhysRevB.94.064501} \\
 & $5\pm1$ & $0.32\pm0.01$ & $100\pm30$ & $12\pm1$ & $0.000\pm0.005$ & $164\pm10$ & 38--290 \cite{PhysRevB.96.140501} \\
 & $8\pm2$ & $0.74\pm0.03$ & $11\pm40$ & $27\pm2$ & $0.000\pm0.007$ & $166\pm9$ & 38--290 \cite{singh2018CaKFe4As4} \\
 & $2.1\pm0.4$ & $0.32\pm0.02$ & $65\pm10$ & $10.1\pm0.8$ & $0.000\pm0.002$ & $178\pm10$ & 38--290 \cite{pyon2019CaKFe4As4} \\
\hline
\multirow{7}{*}{FeSe} & $31\pm7$ & $1.3\pm0.2$ & $742\pm100$ & $32\pm5$ & $0.099\pm0.003$ & $197\pm7$ & 95--200 \cite{PhysRevB.110.224508} \\
 & $4\pm2$ & $0.48\pm0.04$ & $277\pm30$ & $12.1\pm0.5$ & $0.006\pm0.002$ & $200\pm4$ & 95--200 \cite{tanatar2016origin} \\
 & $0.0\pm0.2$ & $0.57\pm0.07$ & $373\pm200$ & $14\pm6$ & $0.00\pm0.03$ & $200\pm30$ & 95--200 \cite{kasahara2014field} \\
 & $6\pm1$ & $0.58\pm0.04$ & $280\pm40$ & $14\pm2$ & $0.000\pm0.007$ & $201\pm10$ & 95--200 \cite{watson2015emergence} \\
 & $5\pm1$ & $0.61\pm0.02$ & $356\pm30$ & $15.5\pm0.3$ & $0.000\pm0.001$ & $198\pm3$ & 95--200 \cite{zhou2021FeSe} (S1) \\
 & $24\pm4$ & $0.66\pm0.02$ & $390\pm30$ & $17.0\pm0.2$ & $0.000\pm0.002$ & $198\pm5$ & 95--200 \cite{zhou2021FeSe} (S2) \\
 & $53\pm4$ & $0.87\pm0.03$ & $545\pm50$ & $23.8\pm0.4$ & $0.000\pm0.004$ & $191\pm4$ & 95--200 \cite{zhou2021FeSe} (S3) \\
\end{tabular}
\end{ruledtabular}
\end{center}
\end{table*}

\section{Temperature dependence of $\sigma_s$}
\label{app_sigmas}

Throughout the main text the saturating channel is taken to be temperature
independent.  To test how much the analysis relies on that assumption, we
repeat it with

\begin{equation}
\sigma_s(T)=\sigma_{s,0}+S\,T ,
\label{eq_sigmas_T}
\end{equation}

fitted simultaneously with $\rho_i$ and with $\rho_0$ fixed exactly as in
Appendix~\ref{app_fit}.

We restrict the drift to a linear term because fitting the more general form
$\sigma_s=\sigma_{s,0}+S\,T^{\beta}$ leaves $\beta$ poorly constrained. Namely, allowing a free exponent does not improve the fit, so we adopt a linear temperature dependence as a simple parametrization of the weak variation of $\sigma_s$ over the measured range.

\begin{figure}
    \includegraphics[width=0.45\textwidth]{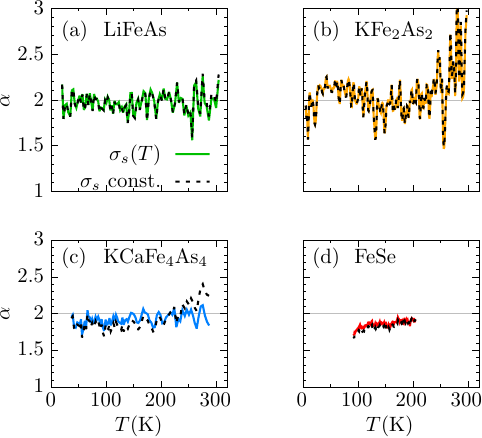}
    \caption{Effect of allowing a temperature-dependent
    saturating conductivity on the local exponent $\alpha(T)$, for
    (a) LiFeAs, (b) KFe$_2$As$_2$, (c) KCaFe$_4$As$_4$ and (d) FeSe, the same representative datasets as Fig.~\ref{fig:app_rhoi}. Dotted black lines use a constant $\sigma_s$, as in the main text; solid coloured lines use $\sigma_s(T)=\sigma_{s,0}+S\,T$ [Eq.~\ref{eq_sigmas_T}], with $\rho_0$ fixed. $\alpha$ is extracted pointwise using Eq.~\ref{eq_rhoi_extract} and the grey line marks $\alpha=2$.}
    \label{fig:sigmas}
\end{figure}

The result is shown in Fig.~\ref{fig:sigmas}. The additional freedom is used only sparingly: across the full fitting window the fitted $\sigma_s$ changes by $\sim3\%$, $\sim1\%$, $\sim12\%$ and $\sim5\%$ for LiFeAs,  KFe$_2$As$_2$, KCaFe$_4$As$_4$ and FeSe respectively, the last within the tetragonal phase window. In every case, $\alpha$ remains close to $2$ [Fig.~\ref{fig:sigmas}].

Equation~\ref{eq_rhoi_extract} shows that the extraction of $\rho_i$
becomes increasingly sensitive to experimental uncertainties and to
the choice of $\sigma_s$ as $\rho-\rho_0$ approaches $\rho_s$.
This amplification may contribute to the increased scatter in $\alpha$
above $\sim250$\,K in cases like KFe$_2$As$_2$.

\section{Estimation of $E_F$}
\label{app-EF}

To estimate the values of Fermi energy used in Fig.~\ref{fig:A2}(a), we consider data from quantum oscillation experiments. 
For a parabolic band of effective mass $m^*$ and Fermi momentum $k_F$, the Shubnikov--de Haas frequency is related to the extremal orbit area $A$ by $F=\hbar A/(2\pi e)$, with $A=\pi k_F^2$. This yields
\begin{equation}
E_F=\frac{e\hbar F}{m^*}.
\label{eq:EF3}
\end{equation}

Equation~(\ref{eq:EF3}) was used to estimate the Fermi energies of LiFeAs, FeSe, and KFe$_2$As$_2$ from the quantum-oscillation data reported in Refs.~\cite{terashima2014anomalous, doi:10.1143/JPSJ.79.053702, PhysRevLett.108.047002}. We compute $E_F$ for each reported pocket, use their average as the characteristic Fermi energy of the compound, and take the corresponding standard deviation as the error bar shown in Fig.~\ref{fig:A2}(a). These error bars should therefore be interpreted as reflecting the spread of Fermi energies among the different pockets, rather than the uncertainty of an individual measurement.

For KCaFe$_4$As$_4$, no quantum-oscillation frequencies were found in the literature; only an estimate of their order of magnitude is available in Ref.~\cite{PhysRevB.97.134501}. In Fig.~\ref{fig:A2}(a), we used $E_F=(50\pm30)$\,meV.

\section{Estimation of $T_{BG}$ and $T^*$ in LiFeAs}
\label{app_estimates}

\subsection{E–ph backscattering scale}

In systems with small cylindrical Fermi surfaces, el-ph backscattering can lead to a $T^2$ resistivity over a limited intermediate temperature range~\cite{PhysRevB.18.1849}.  
The onset of this regime requires that thermally populated acoustic phonons carry momenta of order $q \sim 2k_F$, where $k_F$ is the Fermi momentum.  
This condition defines a lower bound $T_{BG}$ below which phonon backscattering is suppressed, and a steeper $T^5$ behavior is expected. The crossover temperature is estimated as $k_B T_{BG} = 2 \hbar v_s k_F$,
where $v_s$ is the sound velocity.

For LiFeAs, to estimate a lower bound for $T_{BG}$, we take the smallest relevant Fermi momentum $k_F \sim 0.15\,\text{\AA}^{-1}$ from ARPES measurements~\cite{PhysRevLett.105.067002}, and a sound velocity $v_s = 5340\,\mathrm{m/s}$ from thermodynamic data~\cite{abd2011calculated}. Inserting these values, we obtain $T_{BG}\approx120\,\mathrm{K}$.

Thus, below this temperature, e-ph backscattering is expected to be strongly suppressed.  
The observation of a robust $T^2$ resistivity in LiFeAs even below $T_{BG}$, combined with the absence of a crossover to a higher power-law behavior, supports the interpretation that e-e scattering dominates the transport over the entire measured temperature range.

We note that while it could be possible that, at temperatures above $T_{BG}$, e-ph scattering contributes to the $T^2$ trend,  the fact that the prefactor of the $T^2$ resistivity remains essentially constant from low temperature up to room temperature suggests that this contribution is not significant.%

\subsection{Coherence scale}

Combining the low-temperature quadratic coefficient of LiFeAs, $A_2\simeq2.1\times10^{-5}\,\mathrm{m\Omega\,cm/K^2}$ (Table~\ref{table-fits}), with low-temperature quantum-oscillation data for its Fermi-surface pockets allows an approximate estimate of the coherence scale. Assuming spin-degenerate, two-dimensional parabolic pockets with a common transport scattering rate $\Gamma=\hbar/\tau$, the Drude conductivity gives
\begin{equation}
\begin{aligned}
\sigma &= \frac{2e^2}{hc}\frac{\sum_i\varepsilon_{0,i}}{\Gamma}, \\
T^* &= \frac{hc}{2e^2A_2\sum_i(\varepsilon_{0,i}/k_B)},
\end{aligned}
\label{eq:sigma_sum_eps0}
\end{equation}
where $c=6.3151(8)\,\text{\AA}$ and $T^*$ is defined by $\Gamma_{\mathrm{ee}}(T^*)=k_BT^*$.

Using $\varepsilon_{0,i}=e\hbar F_i/m_i^*$, the quantum-oscillation data of Ref.~\cite{PhysRevLett.108.047002} give orbit-based energy scales $\varepsilon_{0,i}/k_B$ of approximately $620$, $356$, and $723$\,K for the $\delta$, $\beta$, and $\epsilon$ orbits, respectively. Under the preferred orbit assignment, the latter two correspond to different extremal sections of the same electron sheet. Averaging these two values and using the first for the other electron sheet gives $\sum_i\varepsilon_{0,i}/k_B\simeq1160$\,K and hence $T^*\simeq34$\,K.

This estimate neglects hole transport and replaces the warped electron sheets by representative cylinders. Within the same common-rate approximation, additional conducting weight would reduce $T^*$; for example, a total effective energy scale of approximately $1900$\,K would give $T^*\simeq20$\,K. Such additional weight is not determined by the measured orbits. We therefore consider $20$--$35$\,K an illustrative range, rather than quantitatively established bounds; the estimate primarily identifies a coherence scale of a few tens of kelvin.

\section{A multiband toy model}
\label{app_toymodel}

Here we use a minimal multiband model to show that the
decomposition $\sigma=\sigma_i+\sigma_s$ is compatible with multiband
transport when the electronic lifetimes are parametrized by a Fermi-liquid
contribution and a finite contribution at elevated temperature.

Using the two-dimensional parabolic-pocket result summarized
in Appendix~\ref{app_estimates}, the conductivity of pocket $j$ satisfies
$\sigma_j\propto\epsilon_{0,j}\tau_j$.

We assume for each pocket the phenomenological crossover form $\tau_j(T)=\frac{a_j}{T^2}+\tau_{s,j}$, where the first term represents the Fermi-liquid contribution and $\tau_{s,j}$ is a short, weakly temperature-dependent lifetime. This simple form also provides a very good parametrization of the inverse scattering rate obtained in the DMFT calculations of Ref.~\cite{PhysRevLett.111.036401}.

Summing over pockets gives

\[
\sigma(T)\propto
\frac{1}{T^2}\sum_j\epsilon_{0,j}a_j
+
\sum_j\epsilon_{0,j}\tau_{s,j}
\equiv \sigma_i(T)+\sigma_s.
\]
Under this lifetime ansatz, the multiband conductivity separates into a $T^{-2}$ term and a temperature-independent contribution.  The effective parameters $A_2$ and $\rho_s$ encode different material-dependent weighted sums over the Fermi-surface pockets.

More generally, allowing $\tau_{s,j}\propto T^{-\beta_j}$ does not alter the qualitative picture provided that $\beta_j<2$, since this contribution then decreases more slowly than the Fermi-liquid term and increasingly dominates the conductivity at elevated temperature.

As emphasized in the main text and in Ref.~\cite{PhysRevLett.111.036401}, generally, even within a quasiparticle description, the resistivity need not saturate when the scattering rate does, due to the temperature dependencies of the transport kernel $K$, which include e.g., quasiparticle residue. In this regard, multiband \textit{ab initio} calculations of the resistivity of LiFeAs and related compounds are of high interest for future work.

\bibliography{mibib}
\end{document}